\documentclass[12pt]{article}
\usepackage[english]{babel}
\usepackage{graphicx,epsfig}
\makeindex \textwidth 170mm \textheight 220mm \topmargin -5mm
\newcommand{\be}{\begin{equation}}
\newcommand{\ee}{\end{equation}}
\newcommand{\bea}{\begin{eqnarray}}
\newcommand{\eea}{\end{eqnarray}}

\def\rfr#1{eq. (\ref{#1})}
\def\rfrs#1#2{eqs. (\ref{#1})-(\ref{#2})}
\def\Rfr#1{Eq. (\ref{#1})}

\def\bb{\bibitem}
\def\eqi{\begin{equation}}
\def\eqf{\end{equation}}
\def\eqia{\begin{eqnarray}}
\def\eqfa{\end{eqnarray}}
\def\btab{\begin{tabular}}
\def\etab{\end{tabular}}
\def\bar{\begin{array}}
\def\ear{\end{array}}

\def\GR{General Relativity}
\def\grl{general relativistic}
\def\wfs{weak--field and slow--motion approximation}
\def\leti{Lense--Thirring}
\def\grc{gravitomagnetic}

\def\se{systematic error}
\def\er{error}
\def\zh{even zonal harmonics}

\def\gp{geopotential}

\def\lg{{\rm LAGEOS}}
\def\lgg{{\rm LAGEOS} II}

\def\lb#1{\label{#1}}
\def\pc{precession}
\def\nd{node}
\def\pg{perigee}
\def\nl{nodal}

\def\sa{semimajor axis}
\def\ec{eccentricity}
\def\ic{inclination}

\def\et{Earth}
\def\ef{effect}
\def\dt#1{\dot{#1}}
\def\mlt{{\rm \mu_{LT}}}
\def\dmu{\frac{\delta\mlt}{\mlt}}
\def\zs{{\rm zonals}}
\def\st{satellite}
\def\lt{_{\rm{LT}}}
\def\rp#1#2{{#1\over#2}}

\begin{document}
\begin{titlepage}
\begin{flushright}
\today\\
BARI-TH/00\\
\end{flushright}
\vspace{.5cm}
\begin{center}
{\LARGE The LARES mission revisited: an alternative scenario}
\vspace{1.0cm}
\quad\\
{Lorenzo Iorio$^{\dag}$\\ \vspace{0.1cm}
\quad\\
{\dag}Dipartimento di Fisica dell' Universit{\`{a}} di Bari, via
Amendola 173, 70126, Bari, Italy\\ \vspace{0.2cm} } \vspace{0.2cm}
\quad\\
{David M. Lucchesi$^{\sharp}$\\ \vspace{0.1cm}
\quad\\
{$\sharp$}Istituto di Fisica dello Spazio Interplanetario,
IFSI/CNR, Via Fosso del Cavaliere n. 100, 00133, Roma, Italy\\
\vspace{0.2cm} } \vspace{0.2cm}
\quad\\
{Ignazio Ciufolini$^{\ddag}$\\ \vspace{0.1cm}
\quad\\
{\ddag}Dipartimento di Ingegneria dell'Innovazione
dell'Universit{\`{a}} di Lecce, via per Arnesano, 73100, Lecce,
Italy \\
\vspace{0.2cm}} \vspace{1.0cm}

{\bf Abstract\\}
\end{center}

{\noindent \small  In the original LARES mission the \grl\
\leti\ \ef\ would be detected by using as observable the sum of
the residuals of the nodes of the existing passive geodetic
laser-ranged LAGEOS \st\ and of its proposed twin LARES. The
proposed nominal orbital configuration of the latter one would
reduce the \se\ due to the mismodelling in the \zh\ of the \gp,
which is the main source of error, to $0.3\%$, according to the
most recent Earth gravity model EGM96. This observable turns out
to be sensitive to possible departures of the LARES orbital
parameters from their nominal values due to the orbital injection
errors. By adopting a suitable combination of the orbital
residuals of the \nd s of LAGEOS, LAGEOS II and LARES and the \pg
s of LAGEOS II and LARES it should be possible to reduce the error
due to the \gp\ by one order of magnitude, according to the EGM96
gravity model. Moreover, the sensitivity to the orbital injection
errors should be greatly reduced. According to a preliminary
estimate of the error budget, the total error of the experiment
should be reduced to less than 1$\%$. In the near future, when the
new data of the terrestrial gravitational field from the CHAMP and
GRACE missions will be available, a further increase in the
accuracy should be obtained. The proposal of placing LARES in a
polar 2,000 km altitude orbit and considering only its nodal rate
would present the drawback that even small departures from the
polar geometry would yield notable errors due to the mismodelled
\zh\ of the \gp, according to the EGM96 model. }
\end{titlepage} \newpage \pagestyle{myheadings} \setcounter{page}{1}
\vspace{0.2cm} \baselineskip 14pt

\setcounter{footnote}{0}
\setlength{\baselineskip}{1.5\baselineskip}
\renewcommand{\theequation}{\mbox{$\arabic{equation}$}}
\noindent

\section{Introduction}
In its \wfs\ \GR\ predicts that, among other things,  the orbit of
a test particle freely falling in the gravitational field of a
central rotating body is affected by the so called \grc\ dragging
of the inertial frames or
\leti\ \ef. More precisely,  the longitude of the ascending \nd\ $\Omega$ and
the argument of the \pg\ $\omega$ of the orbit [{\it Sterne},
1960] undergo tiny \pc s according to [\textit{Lense and
Thirring}, 1918] \eqia \dot\Omega \lt & = &
\frac{2GJ}{c^{2}a^{3}(1-e^{2})^{\frac{3}{2}}},\\
\dot\omega \lt & = &
-\frac{6GJ\cos{i}}{c^{2}a^{3}(1-e^{2})^{\frac{3}{2}}},\eqfa in
which $G$ is the Newtonian gravitational constant, $J$ is the
proper angular momentum of the central body, $c$ is the speed of
light $in\ vacuum$, $a,\ e$ and $i$ are the \sa, the \ec\ and the
\ic, respectively, of the orbit of the test particle.

The first measurement of this \ef\ in the gravitational field of
the \et\ has been obtained by analyzing a suitable combination of
the laser-ranged data to the existing passive geodetic \st s \lg\
and \lgg\ [\textit{Ciufolini et al.,} 1998]. The observable [{\it
Ciufolini}, 1996] is a linear trend with a slope of 60.2
milliarcseconds per year (mas/y in the following) and includes the
residuals of the nodes of \lg\ and \lgg\ and the \pg\ of
\lgg\footnote{The \pg\ of \lg\ was not used because it introduces
large observational errors due to the smallness of the \lg\ \ec\
[{\it Ciufolini}, 1996] which amounts to 0.0045.}. The
\leti\ precessions for the \lg\ satellites amount to \eqia
\dot\Omega\lt^{\rm LAGEOS}& = & 31\ \textrm{mas/y},\\
\dot\Omega\lt^{\rm LAGEOS\ II} & = & 31.5\ \textrm{mas/y},\\
\dot\omega\lt^{\rm LAGEOS} & = & 31.6\ \textrm{mas/y},\\
\dot\omega\lt^{\rm LAGEOS\ II} & = & -57\ \textrm{mas/y}. \eqfa
The total relative accuracy of the measurement of the solve-for
parameter $\mlt$, introduced in order to account for this \grl\
\ef, is about $2-3\times 10^{-1}$ [{\it Ciufolini et al.}, 1998].

In this kind of experiments using Earth satellites the major
source of \se s is represented by the aliasing trends due to the
classical secular precessions [\textit{Kaula}, 1966] of the \nd\
and the \pg\ induced by the mismodelled \zh\ of the \gp\ $J_2,\
J_4,\ J_6,...$ Indeed, according to the present knowledge of the
\et's gravity field based on the EGM96 model [\textit{Lemoine et
al.}, 1998], the $J_{2n}$ errors are comparable in size with the
\grc\ precessions of interest, especially for the first two even
zonal harmonics. In the performed \lg\ experiment the adopted
observable allowed for the cancellation of the static and
dynamical effects of $J_2$ and $J_4$. The remaining higher degree
even zonal harmonics affected the measurement at a $13\%$ level.

In order to achieve a few percent accuracy, in
[\textit{Ciufolini}, 1986] it was proposed to launch a passive
geodetic laser-ranged \st- the former {\rm LAGEOS} III - with the
same orbital parameters of \lg\ apart from its inclination which
should be supplementary to that of \lg.

This orbital configuration would be able to cancel out exactly the
classical \nl\ \pc s, which are proportional to $\cos i$, provided
that the observable to be adopted is the sum of the residuals of
the \nl\ \pc s of {\rm LAGEOS} III and LAGEOS \eqi
\delta\dt\Omega^{{\rm III}}+\delta\dt\Omega^{{\rm
I}}=62\mlt.\lb{lares}\eqf Later on the concept of the mission
slightly changed. The area-to-mass ratio of {\rm LAGEOS} III was
reduced in order to make less relevant the impact of the
non-gravitational perturbations, the total weight of the satellite
was reduced to about 100 kg, i.e. to about 25$\%$ of the weight of
LAGEOS, and the eccentricity was enhanced in order to be able to
perform other \grl\ tests: the LARES was born [\textit{Ciufolini},
1998]. The orbital parameters of \lg, \lgg\ and LARES are in Tab.
1.

\begin{table}[ht!]
\caption{Orbital parameters of \lg, \lgg\ and LARES.} \label{para}
\begin{center}
\begin{tabular}{lllll}
\noalign{\hrule height 1.5pt} Orbital parameter & \lg & \lgg &
LARES\\ \hline
$a$ (km) & 12,270 & 12,163 & 12,270\\
$e$ & 0.0045 & 0.014 & 0.04\\
$i$ (deg) & 110 & 52.65 & 70\\
\noalign{\hrule height 1.5pt}
\end{tabular}
\end{center}
\end{table}
At present the LARES experiment is just at a Phase--A stage and
has not yet been approved by any space agency. Although much
cheaper than other proposed and approved complex space--based
missions, funding is the major obstacle in implementing the LARES
project.

In this paper we investigate the possibility of modifying the
original LARES mission in order to achieve significant
improvements in the reduction of some relevant systematic errors.
The paper is organized as follows. In Section 2 we analyze in
detail the impact of the unavoidable orbital injection \er s in
the orbital parameters of LARES on the \se\ induced by the
mismodelling in \zh\ of the \gp\ according to the most recently
released \et\ gravity model EGM96. Moreover, in Section 3 we
propose an alternative configuration which should be able to
reduce this error by one order of magnitude. It adopts as
observable a suitable combination of the orbital residuals of the
\nd s of \lg, \lgg\ and LARES, and the \pg s of \lgg\ and LARES.
It presents also the important advantage that it is almost
insensitive to the errors in the \ic\ of LARES, contrary to the
original {\rm LAGEOS}/LARES only configuration. A further
observable, based only on the nodes of the three LAGEOS satellites
and of Ajisai, is also presented. Some possible less convincing
implications of placing the LARES in a low-altitude polar orbit
are examined in Section 4. Section 5 is devoted to the
conclusions.
%---------------------------------------------------------------
\section{The impact of the even zonal harmonics of the geopotential on the
original LARES mission} Let us calculate the \se\ induced by the
mismodelling in the even degree zonal coefficients $J_2$,
$J_4$,... of the \gp\ on the sum of the classical precessions of
the \nd s of \lg\ and LARES. It is important to stress that it is
the major source of \se\ and cannot be eliminated in any way. We
will use the covariance matrix of the \et\ gravity field model
EGM96 [{\it Lemoine et al.}, 1998] by summing up in a root sum
square fashion the correlated contributes up to degree $l=20$. The
relative error obtained by using the nominal values of Tab. 1
amounts to \eqi\dmu_{\zs}=3\times 10^{-3}.\lb{1zon}\eqf It is not
equal to zero because we have assumed $e_{{\rm LR}}=0.04$ while
$e_{{\rm {\rm LAGEOS}}}=0.0045$. If it was $e_{{\rm LR}}= e_{{\rm
{\rm LAGEOS}}}$, then the classical \nl\ \pc s would be exactly
equal in value and opposite in sign and would cancel out. Note
that the coefficients with which $\delta\dot\Omega^{\rm III}$ and
$\delta\dot\Omega^{\rm I}$ enter the combination of \rfr{lares} do
not depend on any orbital parameters: they are constant numbers
equal to 1; this combination allows to cancel out all the
classical nodal precessions due to the $J_{2n}$ coefficients,
including those induced by $J_2$ and $J_4$ which are cancelled out
$a\ priori$ in the combination used in the \lg\ experiment [{\it
Ciufolini}, 1996]. They are the most effective in aliasing the
\leti\ precessional rates.

Now we will focus on the sensitivity of $\dmu_{\zs}$ to the
possible orbital injection errors in the orbital parameters of
LARES. For a former analysis see [{\it Casotto et al.}, 1990]. It
is particularly interesting to consider the impact of the errors
in the inclination and the \sa. The ranges of variation for them
have been chosen in a very conservative way in order to take into
account low-precision and low-costs injection scenarios.

\begin{figure}[ht!]
\begin{center}
\includegraphics*[width=13cm,height=10cm]{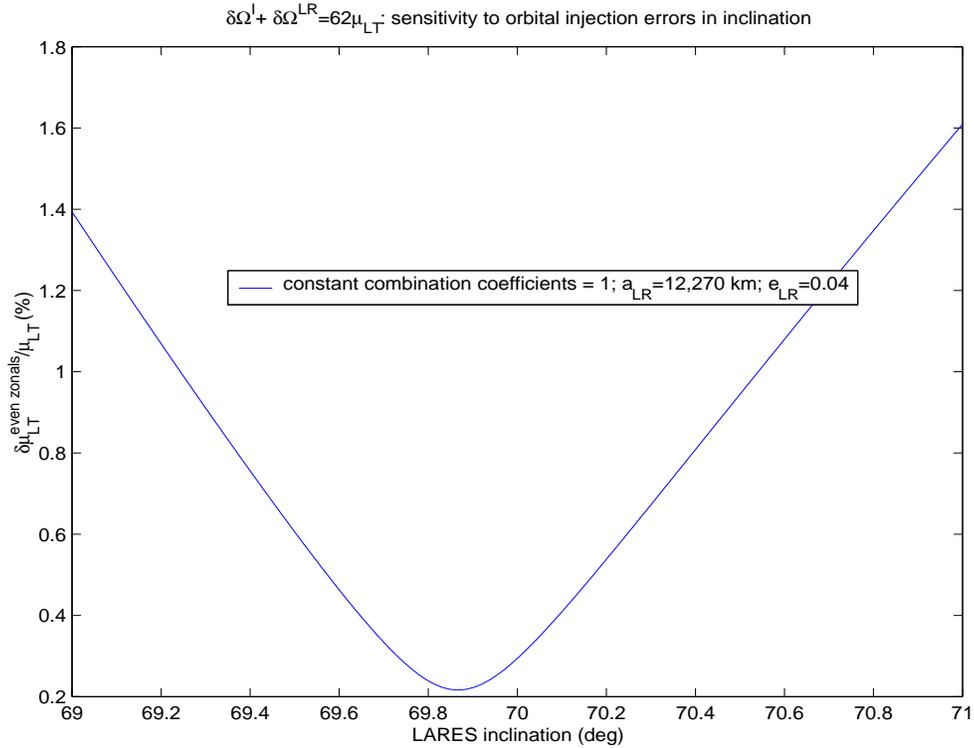}
\end{center}
\caption{\footnotesize Influence of the injection errors in the
LARES inclination on the error due to the even zonal harmonics of
the geopotential.} \label{figura1}
\end{figure}

\begin{figure}[ht!]
\begin{center}
\includegraphics*[width=13cm,height=10cm]{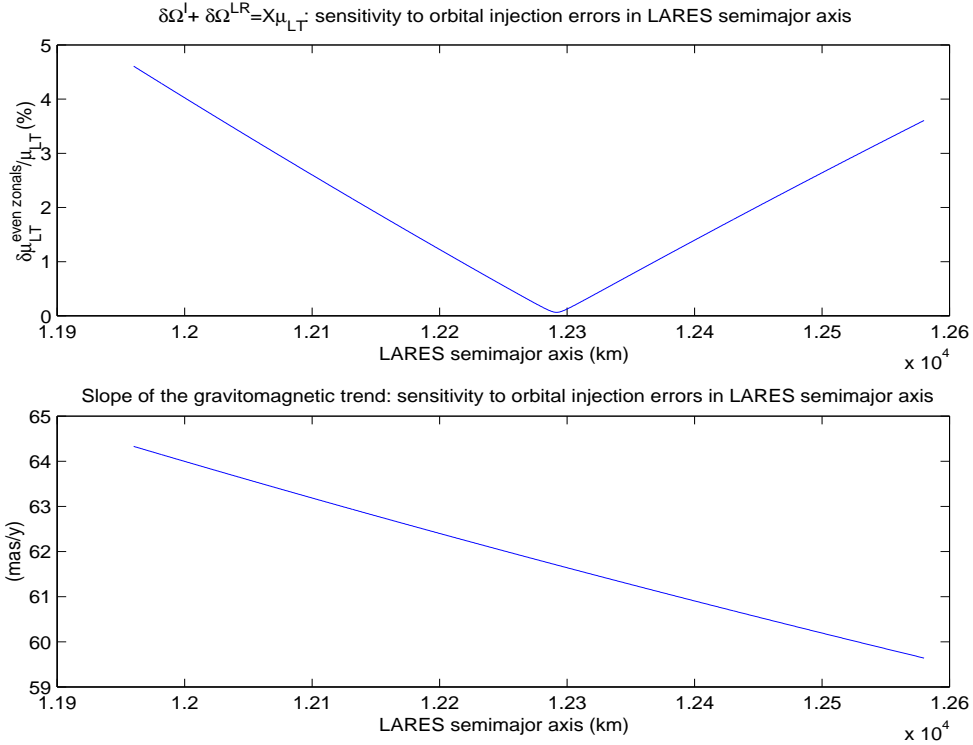}
\end{center}
\caption{\footnotesize Influence of the injection errors in the
LARES \sa\ on the error due to the even zonal harmonics of the
geopotential.} \label{figura2}
\end{figure}

From Fig.1 it is interesting to note that the minimum value of the
systematic zonal error, which amounts to $2.1\times 10^{-3}$, does
not correspond to $i_{{\rm LR}}=70\ \textrm{deg}$ but it is
obtained for a slightly smaller value. It is possible to show that
for $e_{{\rm LR}}=e_{{\rm {\rm LAGEOS}}}$ the minimum is 0 and
that it is attained at $i_{{\rm LR}}=70\ {\rm deg}$. The maximum
error in Fig. 1 amounts to $1.6\times 10^{-2}$. This suggests that
the original LARES project is relatively sensitive to small
departures of $i_{{\rm LR}}$ from its nominal value. Fig. 2 shows
that even more relevant is the sensitivity to the LARES \sa. Also
in this case the minimum is attained at a value of $a_{{\rm LR}}$
smaller than the nominal $a_{{\rm LR}}=12,270$ km. Notice that the
variation of the error is more than one order of magnitude and may
also reach values of some percent. For $e_{{\rm LR}}=e_{{\rm {\rm
LAGEOS}}}$ the minimum error amounts to 0 and it is obtained for
$a_{{\rm LR}}=12,270$ km, as expected. In obtaining Fig. 2 we have
accounted for the dependence of the \nl\
\leti\ \pc\ on $a$ by varying, accordingly, the slope of the \grl\
trend. The sensitivity to eccentricity variations is less
relevant: e.g., by varying it from 0.03 to 0.05 the relative
systematic zonal error increases from $1.6\times 10^{-3}$ to just
$4.6\times 10^{-3}$.
%-------------------------------------------------------------------
\section{An alternative LARES scenario}
Here we will look for an alternative observable involving the
orbital elements of LARES satisfying the following requirements
\begin{itemize}
  \item It should yield a value
for the \se\ due to the mismodelled \zh\ of the \gp\ smaller than
that of the simple sum of the \nd s of \lg\ and LARES. Moreover,
such error should be less sensitive to the departures of the
possible real orbital elements of LARES from the nominal values of
Tab. 1

  \item It should contain and, if possible, reduce,
  the time--varying gravitational and
  non--gravitational part of the error budget
\end{itemize}
These requirements could be implemented by setting up a suitable
orbital combination which cancels out the contributions of as many
mismodelled even zonal harmonics as possible, following the
strategy of the \lg\ experiment outlined in [{\it Ciufolini},
1996]. To this aim we will consider only the \st s of the \lg\
family, both because they are the best laser-ranged targets and
because the gravitational and non-gravitational perturbations
affecting their orbits have been, and will be, extensively and
thoroughly analyzed. Moreover, since they are almost insensitive
to the even zonal harmonics of the geopotential of degree higher
than $l=20$, the use of the covariance matrix of EGM96 up
to\footnote{For higher degrees the reliability of the EGM96 model
is questionable.} $l=20$ should allow for realistic estimates of
the systematic error due to the static part of the terrestrial
gravitational field.

Our result is \eqi \delta\dt\Omega^{\rm {\rm
LAGEOS}}+c_1\delta\dt\Omega^{\rm {\rm LAGEOS}\
II}+c_2\delta\dt\Omega^{\rm LARES}+c_3\delta\dt\omega^{\rm {\rm
LAGEOS} \ II}+c_4\delta\dt\omega^{\rm
LARES}=61.8\mlt,\lb{combi}\eqf with \eqia
c_1 & = & 6\times 10^{-3},\lb{c1}\\
c_2 & = & 9.83\times 10^{-1},\\
c_3 & = & -1\times 10^{-3},\\
c_4 & = & -2\times 10^{-3}\lb{c4}. \eqfa The coefficients $c_i$
given by \rfrs{c1}{c4} have been obtained by solving for the five
unknowns $\delta J_2$, $\delta J_4$ $\delta J_6$, $\delta J_8$ and
$\mu_{\rm LT}$ a nonhomogeneous algebraic linear system of five
equations expressing the observed mismodelled classical
precessions of the nodes of LAGEOS, LARES and LAGEOS II and the
perigees of LAGEOS II  and LARES. They depend on the orbital
parameters of LAGEOS, LAGEOS II and LARES (nominal values of Tab.
1) and are built up so to cancel out all the static and dynamical
contributions of degree $l=2,\ 4,\ 6,\ 8$ and order $m=0$ of the
Earth' s gravitational field.

The relative \se\ due to the $J_{2n},\ n\geq 5$, according to
EGM96 up to degree $l=20$, amounts to \eqi\dmu_{\zs}=2\times
10^{-4},\lb{kazonga}\eqf which is one order of magnitude better
than the result of \rfr{1zon}.

\begin{figure}[ht!]
\begin{center}
\includegraphics*[width=13cm,height=10cm]{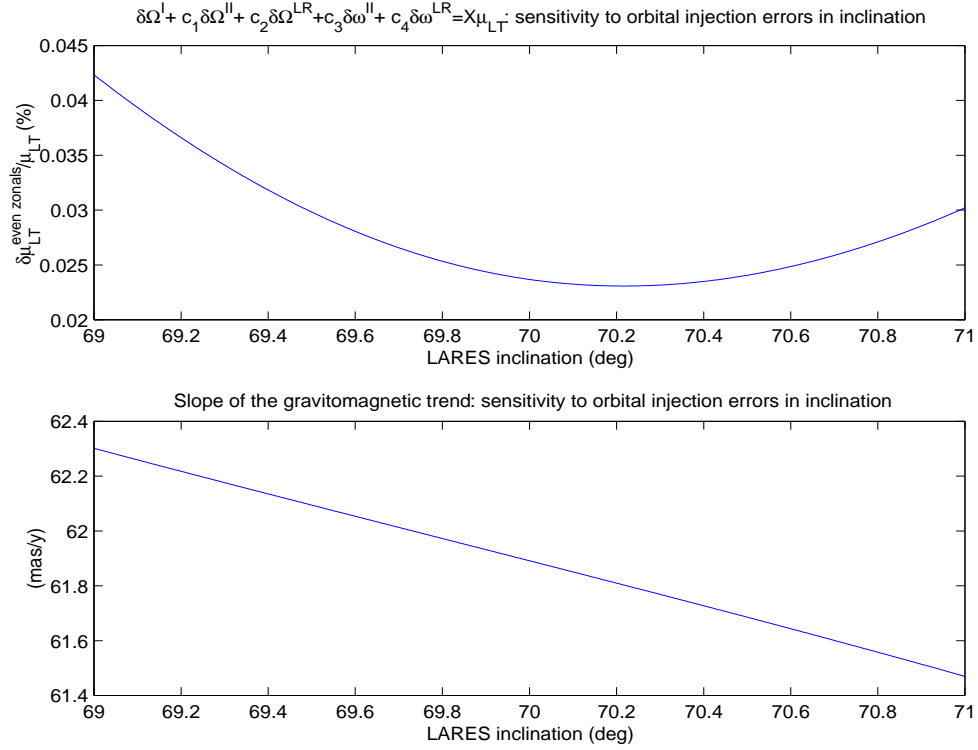}
\end{center}
\caption{\footnotesize Alternative combined residuals: influence
of the injection errors in the LARES \ic\ on the error induced by
the even zonal coefficients of the geopotential.} \label{figura3}
\end{figure}

\begin{figure}[ht!]
\begin{center}
\includegraphics*[width=13cm,height=10cm]{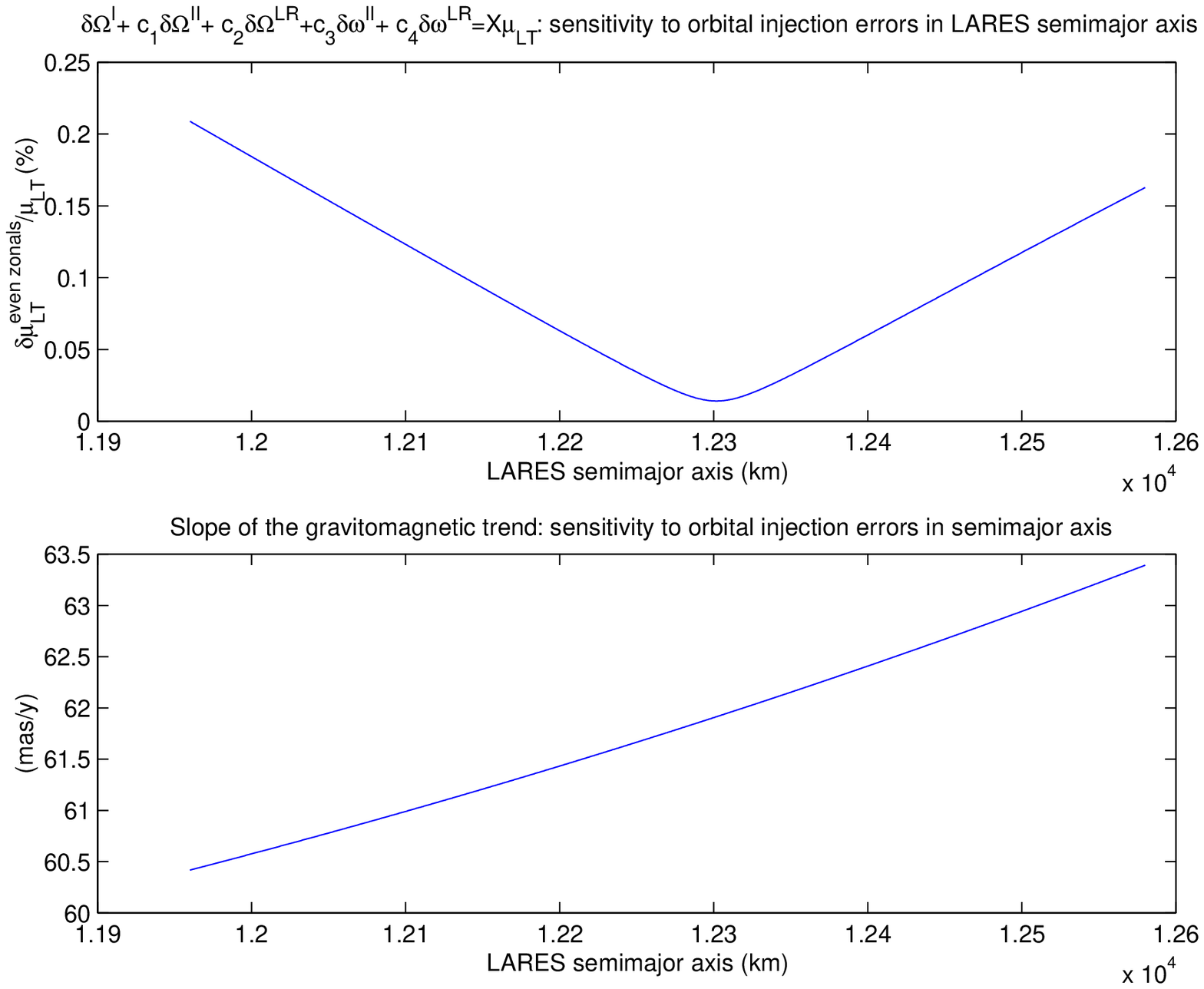}
\end{center}
\caption{\footnotesize Alternative combined residuals: influence
of the injection errors in the LARES \sa\ on the error induced by
the even zonal coefficients of the geopotential.} \label{figura4}
\end{figure}

Fig. 3 and Fig. 4 show the achievements realized in reducing the
sensitivity of the proposed combined residuals to the orbital
injection errors in the LARES orbital elements. In obtaining Fig.
3 and Fig. 4 we have accounted for the dependence on $a_{\rm LR}$
and $i_{\rm LR}$ of both the coefficients and the
\leti\ \pc s: it turns out that the variations in the slope of the
\grl\ trend are very smooth with respect to the nominal value of
61.8 mas/y amounting to few mas/y and the values of the zonal
error are much closer to the nominal one given by \rfr{kazonga}.
Also in this case, the minima are attained at slightly different
values of the LARES orbital elements with respect to the nominal
ones. It is also interesting to note in Fig. 3 that over a 3$\%$
variation of $i_{\rm LARES}$ the error due to the mismodelled
zonal harmonics remain almost constant, while over a 5$\%$
variation of $a_{\rm LARES}$ it changes by 1 order of magnitude,
as it turns out from Fig. 4. However, the result is quite
satisfactory, especially if compared to Fig. 2.
%-------------------------------------------------------------
\subsection{Preliminary error budget estimate}
It is worthwhile noticing that the time-varying gravitational and
non--gravitational orbital perturbations which would affect the
proposed combined residuals would be depressed by the small values
of the coefficients with which some orbital elements enter the
combination.

\begin{itemize}
%-------------------------------------------------------------
\item
{For example, in regard to the Earth solid and ocean tides [{\it
Iorio}, 2001], it is important that the perigees of LAGEOS II and
LARES, which are affected by very long--period uncancelled tidal
perturbations, enter the combination weighted by small
coefficients of order of $10^{-3}$. On the contrary, the tidal
perturbations which would affect the nodes of LAGEOS and LARES,
which enter the combination with coefficients of the order of the
unity, would have periods of short or medium length, so that they
could be averaged out or, at least, could be viewed as empirically
fitted quantities over an observational time span $T_{\rm obs}$ of
few a years only [{\it Iorio and Pavlis}, 2001].}
%--------------------------------------------------------------
\item
{More subtle and complex to model is the action of the
non--gravitational (NG) perturbations. These perturbative effects
depend on the physical and geometrical features of the satellites,
on the geometry of their orbit in space -- orientation and size --
and on the complex interaction of the electromagnetic radiation of
solar and terrestrial origin with the satellites' surfaces.

In particular thermal thrust effects -- due to the uncertainties
of some of their characteristic parameters -- play a crucial role
in the Lense-Thirring effect determination on both the node and
the perigee of LAGEOS--type satellites [${\it Lucchesi}$, 2001;
2002]. These perturbations, such as the solar Yarkovsky-Schach
effect, the Earth Yarkovsky-Rubincam effect and the asymmetric
reflectivity effect, are related to the satellite spin axis
orientation and rate. This opens the problem of the satellites
spin axis vector determination from ground observations in such a
way to establish if the models developed for the spin axis
evolution [{\it Bertotti and Iess}, 1991; {\it Habib et al.,}
1994; {\it Farinella et al.}, 1996; {\it Currie et al.,} 1997;
{\it Bianco et al.,} 2001] are reliable or not. For example, the
previously cited models seem no more able to explain the LAGEOS
spin axis evolution starting from 1997 [{\it $M\acute{e}tris$ et
al.}, 1999]. Moreover, it should also be stressed that such subtle
disturbing accelerations are included in a rather incomplete
manner in the force models of orbit determination softwares like
GEODYN II. It should be underlined that, if and when the LARES
mission will be implemented, the effects of the thermal forces
will be different for the three satellites also because the
dynamical states of their spins will radically differ from each
other. Temporal variations in the reflectivity  coefficients
$C_{\rm R}$ of the LAGEOS satellites have also to be accurately
determined and modelled using the orbital observations. Indeed,
the reflectivity coefficient of the LAGEOS II satellite has been
observed to be changing in time. However, this change can be
accurately modelled from the experimental data, independently of
the Lense--Thirring effect, because of the precise knowledge of
the frequencies and relative amplitudes of the orbital
perturbations due to the direct solar radiation pressure. In our
error budget we have also included this source of error due to the
mismodelling of time variations in the satellites reflectivity
coefficients.

In the following we give the results -- on the LT effect
measurement -- of a numerical simulation and analysis of the
satellites orbit over a 7--year observational time span. We
considered the previously quoted thermal thrust effects, the
direct solar radiation pressure and Earth albedo on the satellites
node and perigee rates. For the definition and characteristics of
these perturbations we refer to literature, while for the most
recent and significative results we refer to {\it $M\acute{e}tris$
et al.}, [1997, 1999] and to ${\it Lucchesi}$ [2001, 2002]. In
Tab. 2, 3 and 4 the nominal results, from the numerical simulation
and analysis, respectively in the case of LAGEOS, LAGEOS II and
LARES are shown. The rates obtained are expressed in mas/y.

%--------------------------------------------------------
\begin{table}[ht!]
\caption{ Non--gravitational perturbations on LAGEOS node and
perigee: nominal values. } \label{tab2}
\begin{center}
\begin{tabular}{lll}
\noalign{\hrule height 1.5pt} Perturbation & $\dot\Omega$ (mas/y)& $\dot\omega$ (mas/y)\\
\hline
Yarkovsky-Rubincam  & 0.2 & 0.07\\
Yarkovsky-Schach  & -0.04 & -77\\
Asymmetric reflectivity  & 6$\times 10^{-4}$ &
52.9 \\
Earth albedo & 1.1 & 145 \\
Direct solar radiation pressure & -7.3 & -40,261 \\
\noalign{\hrule height 1.5pt}
\end{tabular}
\end{center}
\end{table}

%-------------------------------------------------

\begin{table}[ht!]
\caption{ Non--gravitational perturbations on LAGEOS  II node and
perigee: nominal values.} \label{tab3}
\begin{center}
\begin{tabular}{lll}
\noalign{\hrule height 1.5pt} Perturbation & $\dot\Omega$ (mas/y) & $\dot\omega$ (mas/y)\\
\hline
Yarkovsky-Rubincam  & -1.5 & 1 \\
Yarkovsky-Schach  & -0.5 & 150\\
Asymmetric reflectivity  & 3$\times 10^{-3}$ & 152\\
Earth albedo & -1.5 & 57 \\
Direct solar radiation pressure & 36 & -2,693\\
\noalign{\hrule height 1.5pt}
\end{tabular}
\end{center}
\end{table}

%-------------------------------------------------------

\begin{table}[ht!]
\caption{ Non--gravitational perturbations on LARES node and
perigee: nominal values.} \label{tab3}
\begin{center}
\begin{tabular}{lll}
\noalign{\hrule height 1.5pt} Perturbation & $\dot\Omega$ (mas/y) & $\dot\omega$ (mas/y)\\
\hline
Yarkovsky-Rubincam  & -1.1 & 0.3\\
Yarkovsky-Schach  & $-9\times 10^{-4}$ & -2\\
Asymmetric reflectivity  & -- & --\\
Earth albedo & -1 & 24 \\
Direct solar radiation pressure & 8.5 & -1,263\\
\noalign{\hrule height 1.5pt}
\end{tabular}
\end{center}
\end{table}

%------------------------------------------------------
In the simulation performed we neglected the possibility of an
asymmetric reflectivity effect in the case of LARES. This is due
to the particular care with which its structure will be build--up
(see LARES proposal, [{\it Ciufolini}, 1998]) to avoid some of the
problems related with the thermal thrust effects, and also for the
absence of the four Germanium cube--corner retroreflectors, which
could be good candidates to explain at least a part of this
anisotropy in the satellite hemispheres reflectivity. We also
neglected the time variations of the reflectivity coefficients of
the satellites. In the simulation we applied to LARES the spin
model developed for LAGEOS in the {\it Farinella et al.} [1996]
version ({\rm rapid--spin} approximation)\footnote{This model
[{\it Lucchesi}, 2002] still gives very good results in the case
of LAGEOS II, in agreement with the results of {\it Bianco et al.}
[2001] for the satellite rotational rate.}. For LAGEOS we applied
the model starting from 1993, when the gravitational torque gives
the major contribution to the evolution. Of course, as previously
stated, starting from 1997 the model does not give, for the spin
components of LAGEOS, values in good agreement with the
observations.

The larger effects in the analysed orbital elements are, of
course, those due to the direct solar radiation pressure. This is
a rather accurate modelled perturbative effect on passive
satellites and we have assumed a value of about $0.5\%$ for its
uncertainty in the case of LAGEOS--type satellites, limited by the
measurement errors in the determination of the solar constant
$\Phi_{\odot}$ and especially in the knowledge of the satellite
reflectivity coefficient $C_{\rm R}$ and in its temporal
variation. Following a conservative approach we can assume an
uncertainty of about $20\%$ for the other perturbations. This is
probably a pessimistic assumption, particularly for the
perturbative effects of the Earth Yarkovsky--Rubincam effect, but
is also a way to get an {\rm upper}--{\rm bound} {\rm order}--{\rm
of}--{\rm magnitude} estimate of the non--gravitational
perturbations error budget. We can then combine linearly --
following Eq. (9) -- the rates obtained in such a way to estimate
the impact  of the uncertainty of each perturbation in the
Lense--Thirring effect determination. The results obtained are
reported in Tab. 5.

%------------------------------------------------------------------
\begin{table}[ht!]
\caption{Errors on the Lense--Thirring measurement due to the
mismodelled non--gravitational perturbations.} \label{tab5}
\begin{center}
\begin{tabular}{ll}
\noalign{\hrule height 1.5pt} Perturbation & $\rp{\delta \mu_{\rm LT}} {\mu_{\rm LT}}$\\
\hline
Yarkovsky-Rubincam  & $3\times 10^{-3}$\\
Yarkovsky-Schach & $6\times 10^{-4}$  \\
Asymmetric reflectivity & $5\times 10^{-4}$ \\
Earth albedo & $1\times 10^{-4}$ \\
Direct solar radiation pressure & $5\times 10^{-4}$\\
\noalign{\hrule height 1.5pt}
\end{tabular}
\end{center}
\end{table}
%------------------------------------------------------------------
As we can see, the largest effect is that of the Earth
Yarkovsky--Rubincam perturbation. This is due to the fact that all
the perigee rate effects are depressed by the very small values of
the coefficients $c_3$ and $c_4$, then the major contributions in
the combination are due to the nodes of LAGEOS and LARES.
%Indeed,
%this result is obvious because of the supplementary orbital
%configuration of these satellites.
It is also important to stress that, while the asymmetric
reflectivity and solar Yarkovsky perturbations gives only
long--term periodic effects in the analysed elements, those due to
the Earth Rubincam perturbation are both secular and periodic in
the nodal and perigee rates. Of course, when the resulting
time--varying perturbations exhibit harmonic behavior with known
relatively short periods, over an observational time span $T_{\rm
obs}$ of some years, they could be fitted and removed from the
signal -- as for the tidal perturbations -- with an improvement in
the
 {\rm root--mean--square} of the residuals. But this technique cannot be applied
in the case of the secular non--gravitational perturbations in the
satellites node and perigee.

To determine the error budget estimate of the Lense--Thirring
effect measurement due to the analysed non--gravitational
perturbations we have conservatively added their contributions. We
finally obtain

\eqi\rp{\delta \mu_{\rm LT}}{\mu_{\rm LT}}_{\rm NG}\sim 5\times
10^{-3}.\lb{EB}\eqf

As we can see, the impact of the mismodelled non--gravitational
perturbations uncertainties is below $1\%$ of the relativistic
parameter $\mu_{\rm LT}$, even with our conservative approach.
Notice also that it is larger than the error by the static part of
the geopotential.

%-------------------------------------
\noindent We finally conclude this section underlining a few
points

\item
{It should be noticed that, according to certain pessimistic
evaluations, the true, realistic errors in the perigees rates of
LAGEOS--like satellites induced by various sources of systematic
biases might even amount to 100$\%$ of their Lense--Thirring
shifts. However, even in this case, the impact on our proposed
configuration would amount to 2$\times 10^{-3}$ thanks to the
small coefficients with which the perigees of LAGEOS II and LARES
enter the proposed combination.}
%---------------------------------
\item
{Moreover, the observational error in the {\rm LAGEOS} II and
LARES perigees, which are undoubtedly difficult to measure for low
eccentric \st s as \lg\ due to the small value of their
eccentricity, would have an impact of the order of $1\times
10^{-4}$ by assuming an uncertainty of the order of 1 {cm} over 1
year in the \st's position.}
%-----------------------------
\item
{Finally, preliminary estimates of the standard statistical error
in the solve-for least square parameter $\mu_{\rm LT}$, based on
simulations encompassing the present models of the time-dependent
LAGEOS perturbations [{\it Iorio}, 2001; {\it Lucchesi,} 2001;
2002] and the noise level reported in the Lense-Thirring LAGEOS
experiment, yield a value of the order of $10^{-3}$.}
%---------------------------------------------------
}\end{itemize}

So, it should not be unrealistic to predict a total uncertainty
below $1\%$, according to the present--day force models.
%------------------------------------------------------
\subsection{A nodes-only combination}
In order to avoid the use of the perigees, which, independently of
the coefficients with which they would enter the observable, are
more sensitive than the nodes to a large set of classical
gravitational and non--gravitational perturbations, the following
alternative combination may be proposed as well\eqi
\delta\dt\Omega^{\rm {\rm LARES}}+k_1\delta\dt\Omega^{\rm {\rm
LAGEOS}}+k_2\delta\dt\Omega^{\rm Ajisai}+k_3\delta\dt\Omega^{\rm
{\rm LAGEOS\ II} }=62.7\mlt,\lb{combii}\eqf with \eqia
k_1 & = & 1.01,\lb{cc1}\\
k_2 & = & 4\times 10^{-5},\\
k_3 & = & 3\times 10^{-3}. \eqfa Also in this case, the
coefficients $k_1,\ k_2$ and $k_3$ have been calculated for the
nominal values of the orbital parameters of LARES. \Rfr{combii}
uses only the nodes of the three LAGEOS satellites and of Ajisai
[{\it Iorio}, 2002] and cancels out the mismodelled contributions
of $J_2,\ J_4$ and $J_6$. The relative error due to the remaining
zonal harmonics of the geopotential would amount to $3\times
10^{-4}$, as in the case of the previously proposed combination
including the perigees of LAGEOS II and LARES. Moreover, the
time--dependent part of the error budget would be mainly dominated
by the nodes of LAGEOS and LARES. We notice that the Ajisai
satellite is much more sensitive than the LAGEOS satellites to the
non--gravitational perturbations. Indeed, its area--to--mass ratio
is larger than that of the LAGEOS satellites. Nevertheless, the
coefficient $k_2$ weighting the Ajisai contribution in the
combination \rfr{combii} is just $4\times 10^{-5}$, i.e. much
smaller than the other weighting coefficients in \rfr{combii}.

As shown in Fig. 5, the main drawback of \rfr{combii} would be its
sensitivity to the orbital injection errors in the LARES
inclination, contrary to \rfr{combi} and Fig. 3. Instead,
regarding the semimajor axis of LARES, \rfr{combii} would be
rather insensitive to its injection error.

\begin{figure}[ht!]
\begin{center}
\includegraphics*[width=13cm,height=10cm]{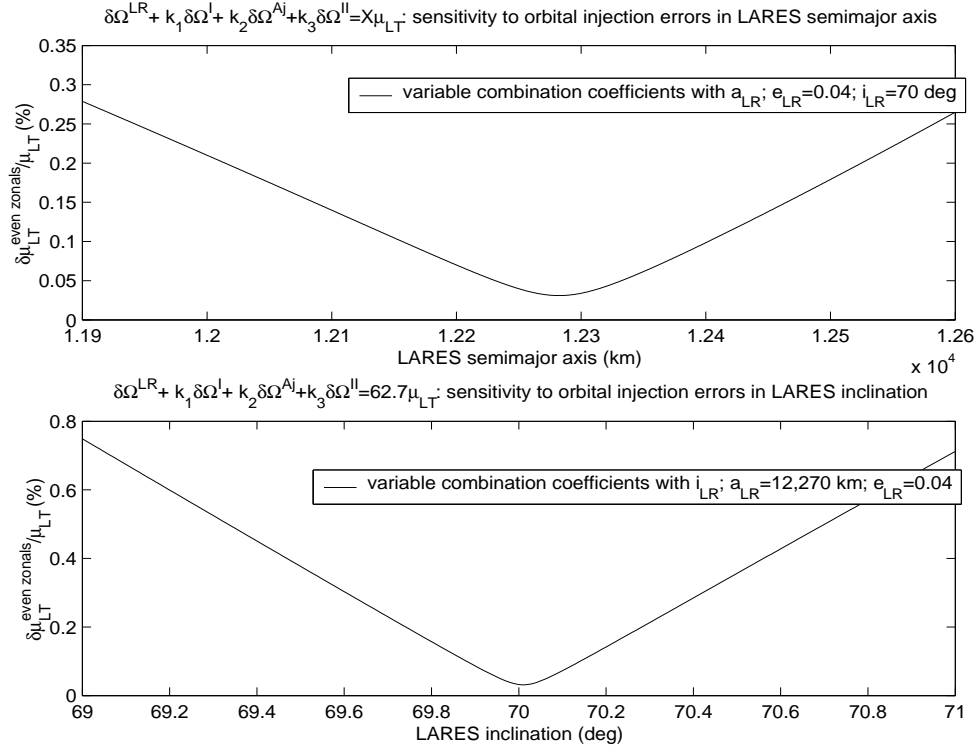}
\end{center}
\caption{\footnotesize Influence of the injection errors in the
LARES \ic\ on the error in the nodes-only combination due to the
even zonal harmonics of the geopotential.} \label{figuray}
\end{figure}
%----------------------------------------------------------------------
\section{The POLARES}
In order to cope with practical launching costs it is currently
under consideration the possibility of inserting the new {\rm
LAGEOS}-like \st\ in a low altitude polar orbit with $i=90$ deg
and $a=8,378$ km obtaining the so called POLARES [{\it Lucchesi
and Paolozzi}, 2001]. The Lense--Thirring shift of its node would
amount to 96.9 mas/y.

If we could obtain and keep exactly an inclination of $90$ deg we
would be able to use the POLARES node only because the classical
mismodelled nodal precessions, which depend on $\cos i$, would
vanish. However, the unavoidable injection errors in the POLARES
inclination, in this case, would be greatly  enhanced by the too
low altitude in the sense that the \se\ due to the \zh\ would blow
up even for relatively small departures from the nominal values.

This is clearly shown by Fig. 6. Also in this case, this result
has been obtained by adding in a root--sum--square fashion the
correlated mismodelled classical nodal precessions with the EGM96
model up to degree $l=20$. Moreover, it should be considered that,
in this case, also the even zonal harmonics of degree higher than
20 would affect the systematic geopotential error.

As expected, for $i_{\rm PL}=90$ deg the systematic zonal error
vanishes.
\begin{figure}[ht!]
\begin{center}
\includegraphics*[width=13cm,height=10cm]{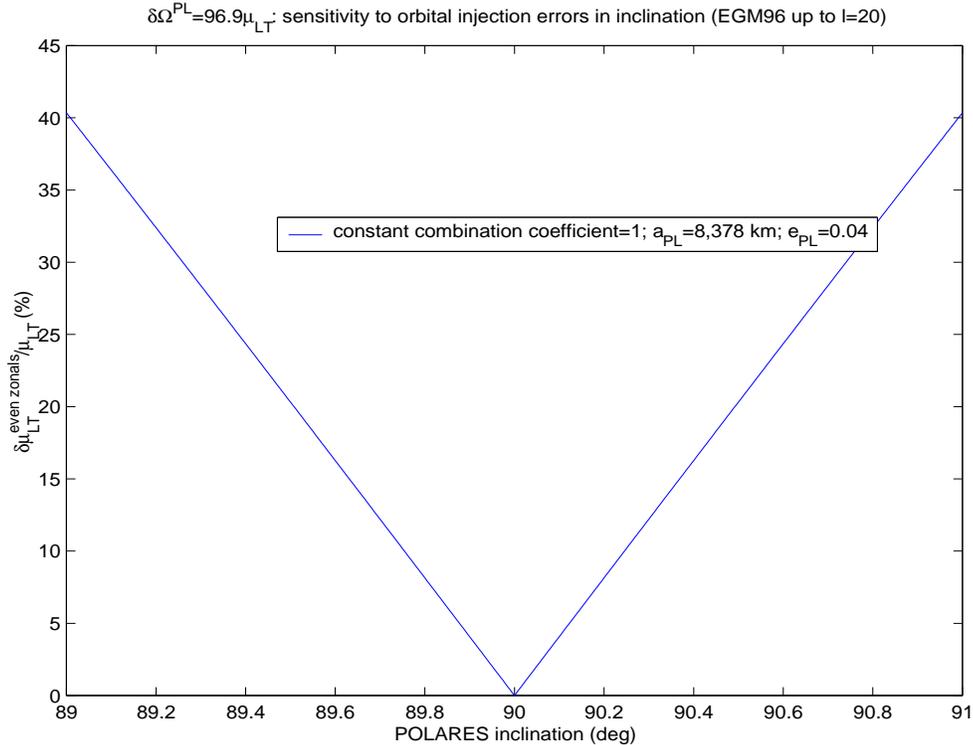}
\end{center}
\caption{\footnotesize Influence of the injection errors in the
POLARES \ic\ on the error in the node only induced by the even
zonal coefficients of the geopotential.} \label{figura3}
\end{figure}
It turns out that even by including the POLARES in some
combinations, this configuration would remain unfavorable.

However, the forthcoming new gravity models from CHAMP and GRACE
will improve the error budget.
%-------------------------------------------------------
\section{Conclusions}
If analyzed from the point of view of the impact of the \se\
induced by the mismodelling in the \zh\ of the \gp, which is the
most important source of systematic error, the originally proposed
LARES observable, consisting of the sum of the nodes of LAGEOS and
LARES, is somehow sensitive to the possible departures of the
original LARES orbital parameters from their nominal values due to
orbital injection errors. The related systematic error could even
raise to few percent, especially as far as the semimajor axis is
concerned. It should be also considered that LARES could be put
into an orbit with a low-cost launcher which, inevitably, would
induce relatively large injection errors. Indeed, the most
expensive part of the implementation of the LARES mission would
just be the launch and in such experiment, which would get a
measurement of the Lense--Thirring effect with accuracy comparable
to the Stanford GP-B mission [{\it Everitt et al.,} 2001], this
could be a serious drawback.

The adoption of the alternative combined residuals proposed here,
including also the \nd\  of LAGEOS II and the \pg s of LAGEOS II
and LARES, would reduce by about one order of magnitude the \se\
due to the \zh\ of the \gp\, decreasing from $0.3\%$ to $0.02\%$,
according to the present--day EGM96 gravity model, and would
greatly reduce the sensitivity of such result to errors in the
LARES orbital parameters. This would yield to less stringent
requirements on the quality and the costs of the launcher to be
adopted.

Preliminary estimates of the error budget, based on the
present--day force models such as the EGM96 Earth gravity model,
show that it would be possible to obtain a total error $\leq 1\%$.
It is very important to notice that when the new data on the
terrestrial gravitational field from the CHAMP and GRACE missions
will be available, the systematic error due to the even zonal
spherical harmonic coefficients of the geopotential will greatly
reduce. The impact of the errors related to the quality of laser
data will further reduce in the near future as well. However, a
careful analysis of the error induced by the spin--dependent,
non--gravitational thermal forces will be required together with
an accurate modelling of the temporal variations of the satellites
reflectivity coefficients. In particular, with the despinning of
the satellites rotational period, more refined thermal models are
needed in such a way to consider the equatorial component of the
perturbing acceleration jointly with the component along the
spin--axis direction, the only one necessary in the rapid--spin
approximation [{\it Vokrouhlichk$\acute{y}$ and Farinella}, 1997].
This perturbative acceleration arises when the longitudinal
gradient of temperature, induced by the slowed down rotation of
the LAGEOS satellites, become significative. It is worth noticing
that, with the proposed observable, the impact of the
non--gravitational orbital perturbations is more relevant than
that of the gravitational perturbations; this fact will be further
enforced when the gravity models from CHAMP and GRACE will be
available.

It should also be possible to adopt a combination involving only
the nodes of LARES, LAGEOS, Ajisai and LAGEOS II. The \se\ due to
the \zh\ of the \gp\ would be equal to the previous case in which
the node of Ajisai would be substituted by the perigees of LARES
and LAGEOS II, but this combination turns out to be more sensitive
to the orbital injection errors in the inclination of LARES.

The possibility of injecting LARES in a low polar orbit at 2,000
{km} of altitude in order to consider only its nodal rate would
be, at the present level of knowledge of the Earth's gravity
field, a not entirely satisfactory solution because, according to
our evaluations based on EGM96, even small deviations from the
projected inclination would lead to an error due to the \zh\ of
the \gp\ of several percents. Anyway, this solution may be
considered as a lower--cost approach to the Lense--Thirring effect
determination in such a way to reduce the error budget below the
$\sim 20\%$ uncertainty of the actual measurement obtained with
LAGEOS and LAGEOS II [{\it Lucchesi}, 2002]. However, the
systematic gravitational errors using a polar satellites will be
reduced with the new gravity models from the CHAMP and GRACE
missions.

The approach outlined here could also be useful for other precise
general relativistic tests, as sketched in [{\it Iorio}, 2002] for
the measurement of the gravitoelectric perigee advance. Moreover,
in view of an ''opportunistic'' approach to general relativistic
measurements, this strategy could easily be  exploited in
encompassing other satellites, proposed or planned to be launched
in the near future for other scopes, if they will be useful for
testing General Relativity as well.

%-----------------------------------------
\section*{Acknowledgements}
L.I. is grateful to L. Guerriero for his support while at Bari.
%-----------------------------------------

\end{document}